# Nature of the superconductor–insulator transition in disordered superconductors

Yonatan Dubi[1], Yigal Meir[1,2] & Yshai Avishai[1,2]

[1]Department of Physics, Ben Gurion University, [2]The Ilse Katz Center for Meso- and Nano-Scale Science and Technology, Ben Gurion University, Beer Sheva 84105, Israel.

**The interplay of superconductivity and disorder has intrigued scientists for several decades. Disorder is expected to enhance the electrical resistance of a system, whereas superconductivity is associated with a zero-resistance state. Although, superconductivity has been predicted to persist even in the presence of disorder[1], experiments performed on thin films have demonstrated a transition from a superconducting to an insulating state with increasing disorder or magnetic field[2]. The nature of this transition is still under debate, and the subject has become even more relevant with the realization that high-transition-temperature (high-$T_c$) superconductors are intrinsically disordered[3–5]. Here we present numerical simulations of the superconductor–insulator transition in two-dimensional disordered superconductors, starting from a microscopic description that includes thermal phase fluctuations. We demonstrate explicitly that disorder leads to the formation of islands where the superconducting order is high. For weak disorder, or high electron density, increasing the magnetic field results in the eventual vanishing of the amplitude of the superconducting order parameter, thereby forming an insulating state. On the other hand, at lower electron densities or higher disorder, increasing the magnetic field suppresses the correlations between the phases of the superconducting order parameter in different islands, giving rise to a different type of superconductor–insulator transition. One of the important predictions of this work is that, in the regime of high disorder, there are still superconducting islands in the sample, even on the insulating side of the transition. This result, which is consistent with experiments[6,7], explains the recently observed huge magneto-resistance peak in disordered thin films[8–10] and may be relevant to the observation of 'pseudogap' phenomena in underdoped high-$T_c$ superconductors[11,12].**

Superconductivity—the occurrence of the zero-resistance state—has been a central issue in solid-state physics for nearly a hundred years. About half a century after its discovery Bardeen, Cooper and Schreiffer[13] (BCS) explained its microscopic foundation. BCS theory attributes superconductivity to pairing of electrons (Cooper pairs), thus creating a many-body coherent macroscopic wavefunction. Electron pairing defines a global order parameter $\Delta$, characterized by its amplitude and phase. According to BCS theory, the suppression of $\Delta$ to zero by increasing temperature $T$ or magnetic field $B$ destroys the superconducting state.

Soon after the emergence of BCS theory, Anderson[1] showed that weak disorder cannot lead to the destruction of pair correlations. Later, Lee and Ma[14] argued that strong disorder gives rise to spatial fluctuations of $\Delta$ along with its suppression in comparison with its value for the clean system, leading eventually to the destruction of the superconducting state. Such a superconductor–insulator transition (SIT) has indeed been observed in disordered thin superconducting films[2]. A similar, magnetic-field-driven SIT has also been observed. This transition has provoked vast interest, and phenomenological theories, valid near the transition, have been put forward[15,16].

Here, on the basis of the first numerical investigation of the SIT starting from a purely microscopic model, the following physical scenario has emerged. In the presence of disorder, the local superconducting order parameter $\Delta(r)$ develops strong spatial fluctuations[14,17,18], such that regions of space where the amplitude of $\Delta$ is large (called 'superconducting islands') are surrounded by regions with relatively small $\Delta$. The system behaves as a bulk superconductor as long as $\Delta$ is different from zero, and the phases of $\Delta(r)$ on two sides of the sample are correlated. Such correlations are established by coherent tunnelling of Cooper pairs between the islands. For weak disorder, increasing $T$ or $B$ suppresses $\Delta$ in the entire sample, before the system loses phase rigidity, and thus superconductivity is destroyed in a way similar to BCS theory. On the other hand, for stronger disorder, increasing $T$ or $B$ leads to the breakdown of phase-coherent paths between the edges of the sample, thereby driving a transition to an insulating state, even when the superconducting order parameter is still finite. The persistence of superconducting correlations in the insulating phase should have far-reaching observable physical consequences.

Our starting point is the microscopic two-dimensional disordered negative-$U$ Hubbard model (see Methods). This model describes electrons propagating on a two-dimensional disordered square lattice, subject to mutual attraction when two electrons, with opposite spin projections, occupy the same site. The model is known to generate a superconducting ground state when no disorder is present. We first demonstrate the formation and evolution of superconducting islands by solving the Bogoliubov–de Gennes[19] equations (described in the Methods) in the presence of both disorder and magnetic field. A topographic colour plot of the spatial distribution of $|\Delta(r)|$, the amplitude of $\Delta$, for a given disorder realization and a finite $B$ is shown in Fig. 1a. The fluctuations in $|\Delta|$ are clearly visible, and one can resolve regions of high $|\Delta|$ surrounded by regions of low $|\Delta|$. However, the Bogoliubov–de Gennes mean-field approach neglects phase fluctuations altogether, and all regions with non-vanishing $\Delta$ are thus phase-correlated. Consequently, within this approximation, as long as $\langle|\Delta|\rangle$—the spatially averaged $|\Delta|$—fails to vanish, the system behaves as a bulk superconductor. With increasing magnetic field, disorder or temperature, there will be a critical point where $\langle|\Delta|\rangle$ vanishes, and the system loses its superconducting nature. Although such a BCS transition is indeed applicable for weakly disordered systems, we show below that this description breaks down for higher disorder, where phase fluctuations play a crucial role.

To take into account phase fluctuations, here we use a newly developed method[12] (see Methods). While neglecting quantum fluctuations, the method allows calculation of thermal averages of phase correlations, thus going beyond the lowest-energy, saddle-point Bogoliubov–de Gennes solution. In Fig. 1b we plot the magnetic-field dependence of the thermally averaged phase correlations $\langle\cos(\delta\theta_i - \delta\theta_j)\rangle$, where $\delta\theta_i$ is the change of phase of $\Delta(\mathbf{r}_i)$ from its mean-field value, and $\mathbf{r}_i$ and $\mathbf{r}_j$ are different points in the sample, indicated by arrows in Fig. 1a. For the points connected by the green arrow in Fig. 1a, the phase correlations hardly change with $B$ (green curve in Fig. 1b), indicating that these points belong to a coherent superconducting island. However, the points connected by blue and red arrows in Fig. 1a lose their phase coherence with increasing $B$. Thus, at this field the coherent macroscopic superconducting system separates into phase-uncorrelated superconducting islands.

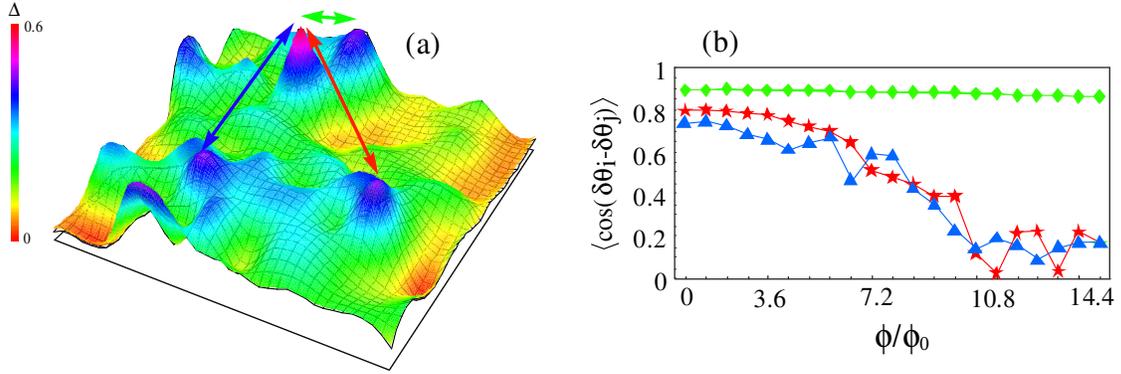

**Figure 1 Spatial fluctuations of the order parameter amplitude and corresponding phase correlations. a**, Spatial distribution of |$\Delta$| for perpendicular field $\phi/\phi_0 = 2.6$ and temperature $T = 0.008t$. The system is of size $12 \times 12$ and has an average electron density $\langle n \rangle = 0.92$, with disorder strength $W/t = 1$. Arrows indicate pairs of points in the sample between which the phase correlations were calculated, shown in **b**. With increasing magnetic field the system separates into islands, the phase correlations between them are suppressed (red and blue arrows and stars and triangles), while for points on the same island (green arrow and diamonds) the phases remain correlated.

Using the same method we demonstrate the emergence of a magnetic-field-driven SIT. In Fig. 2 we plot the spatial average of |$\Delta(\mathbf{r})$| (blue triangles) and the phase correlations (red squares) between the two edges of a superconducting film as a function of $B$. For weak disorder near half filling (Fig. 2a), the superconducting order parameter vanishes at a critical field. Phase correlations between the two sides of the sample persist until that field is reached. On the other hand, at higher disorder (Fig. 2b) or at lower electron density (which corresponds to effective high disorder, inset of Fig. 2a), the critical field $B_c$ is determined by the loss of phase correlations. The amplitude of the order parameter exhibits no particular feature at the transition, and vanishes at a much higher field. Hence, the nature of this transition is entirely distinct from that at low (or no) disorder (and is probably related to the disordered X–Y model[15]). Above $B_c$ the system displays insulating behaviour, but nevertheless supports superconducting correlations, as long as $B$ is lower than the BCS critical field.

Suppression of phase coherence between the superconducting islands with increasing $B$ is displayed in Fig. 3a–c, where on top of the spatial distribution of |$\Delta(\mathbf{r})$| we depict the phase correlation of each point on the lattice with the three points of highest |$\Delta$|—the same points as in

Fig. 1a. Each colour—red, green, blue—indicates correlation with a different point, so that black (mixture of red, green and blue) corresponds to correlation with all points, and white indicates correlations with none. For zero $B$, most points are phase-correlated, but as $B$ increases the islands begin to disconnect, eventually becoming well separated. At such fields the system behaves as an insulator, but both unpaired electrons and Cooper pairs coexist and contribute to the transport process. The persistence of pair correlations beyond the SIT accounts for additional experimental findings, such as local superconducting behaviour on the insulating part of the transition[4–7], and the huge magneto-resistance peak observed in these systems[7–9], which was explained by the competition between contributions of Cooper pairs and unpaired electrons[21].

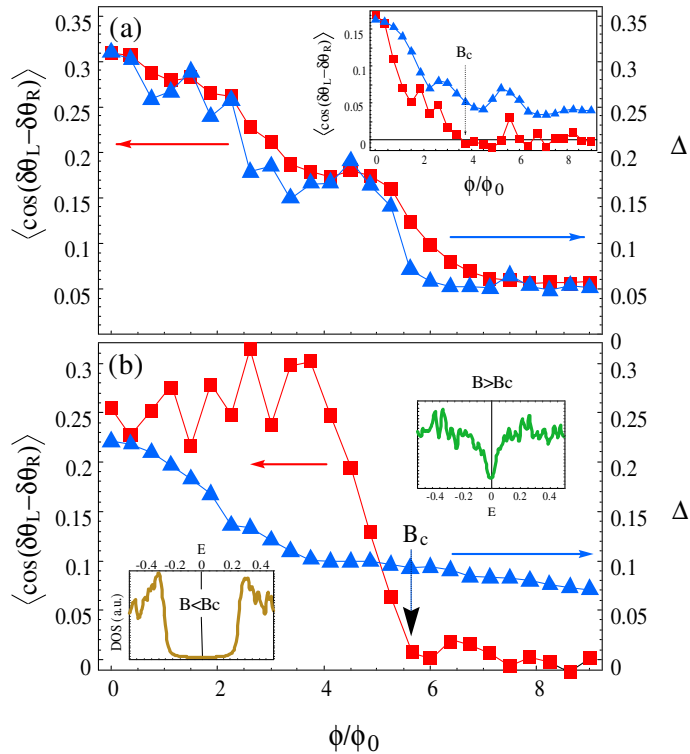

**Figure 2 The superconductor–insulator phase transition with amplitude vanishing and loss of phase coherence. a**, Superconducting order parameter amplitude $|\Delta|$ (blue triangles) and phase correlations between the edges ($\theta_{L(R)}$ stands for order-parameter phases on sites which lie on the left (right) edge of the sample) of a sample of size $15 \times 5$ (red squares), as a function of magnetic field for a weakly disordered sample ($W/t = 0.1$) at electron density $\langle n \rangle = 0.92$ and temperature $T/t = 0.04$. Both $|\Delta|$ and the phase correlations vanish at the same $B$. **b**, The same for a system with stronger disorder ($W/t = 1$), or lower density, $\langle n \rangle = 0.42$ (inset of **a**). Here the phase correlations vanish long before the amplitude. The insets in **b** show the density of states (DOS) at zero field (brown) and on the insulating side of the transition (green), displaying a pseudo-gap feature similar to that observed in high-$T_c$ superconductors.

Page5 of 14

Because local measurements of phase correlations are highly daunting, we propose that the position of the islands and their extent may be experimentally detected by inspecting the dependence of the amplitude of $\Delta(\mathbf{r})$ on a parallel magnetic field $h_\parallel$ that couples only to the electron spin. For clean systems, it is well-known[22,23] that such a field leads to an abrupt vanishing of $\Delta$ and the destruction of the superconducting state into a spin-polarized state, when the gain in Zeeman energy overcomes the superconducting gap. By solving the Bogoliubov–de Gennes equations in the presence of such parallel field, we verify that in the absence of a perpendicular field (when all phases are correlated), superconducting is indeed destroyed abruptly (purple curve in Fig. 3d). However, for higher perpendicular field (thus decreasing the correlations between the phases of the superconducting islands) we find that $|\Delta|$ vanishes in a step-like manner with $h_\parallel$ (blue curve in Fig. 3d), and each step corresponds to the destruction of a different superconducting island. This is depicted in Fig. 3e–h, where the spatial distribution of $|\Delta|$ is plotted for different values of $h_\parallel$. The arrows indicate spatial regions where superconducting vanishes at that field. In Fig. 3i we re-plot the amplitude map, in which each point is now coloured according to the field $h_\parallel$ at which the local $|\Delta(\mathbf{r})|$ has changed. Comparison with Fig. 3c shows that these regions indeed correspond to the superconducting islands as defined by phase correlations and thus they are directly amenable to local experimental probes.

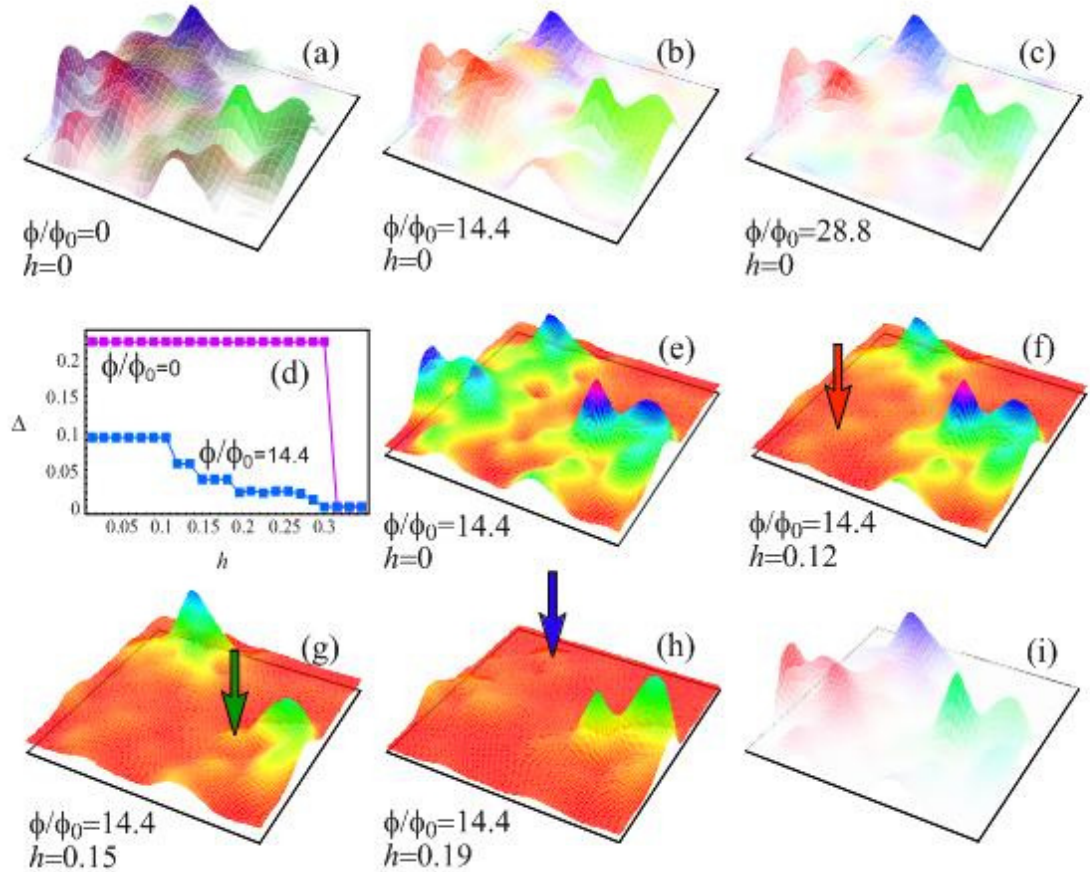

**Figure 3 Superconducting islands observed by phase correlations and by application of a parallel field. a–c**, A spatial map of the phase-correlations: the red, green and blue components of the colour of each point in the sample is proportional to the magnitude of its phase correlations with the three peaks of maximal amplitude (Fig. 1), for different perpendicular magnetic fields, displayed on top of the spatial distribution of the |Δ|. At zero field, phase correlations are long-range, but as the magnetic field is increased the system separates into islands with no inter-island correlations. **d**, ⟨|Δ|⟩ as a function of parallel field, for two different values of the perpendicular magnetic field. At zero perpendicular field the superconducting amplitude vanishes abruptly, while at a finite perpendicular field, ⟨|Δ|⟩ decreases in a series of steps, each island at a time. This is demonstrated in **e–h**, depicting spatial distribution of |Δ| for different values of the parallel field. Arrows indicate the position of the superconducting islands, whose |Δ| vanished at that particular field. **i**, The same distribution of |Δ|, where now each point is coloured by the value of the field at which the amplitude of Δ at that point was suppressed (see **e–h**). Comparing with **c** demonstrates that the islands defined this way are identical to those defined by the loss of phase correlations.

In our model, we have considered only thermal phase fluctuations (owing to computational constraints). That SIT may be explained in terms of thermal fluctuations accounts for many experimental observations in which the universality of a quantum phase transition is not observed, such as the lack of a universal resistance at the transition[24], temperature dependence of the crossing point[25] and the classical X-Y critical exponent[26], and even for the percolation-like behaviour found in some experiments[27–29]. However, it may well be that a similar loss of phase correlations will be driven by quantum fluctuations at low enough temperatures. In fact, recent experiments[7,25] that have explored the competition between thermal and quantum fluctuations (for example, by looking at the dependence on temperature of the crossing point in the resistance–magnetic field plane) demonstrate a continuous crossover from a thermal-fluctuations-driven transition at high temperatures to a quantum-fluctuations-driven transition at low temperatures, the phenomenology of the transition in the two regimes being almost indistinguishable. The non-universality of the critical resistance at the transition may be due to the fact that the dirty boson model for the quantum phase transition does not include the contribution of the unpaired fermions, rather than indicating the irrelevance of the quantum phase-transition scenario.

Finally, while the calculations described above were performed for s-wave superconductors, phase fluctuations have been suggested to be relevant also for high-$T_c$ superconductors[30]. A similar method was recently used[12] to study the phase diagram of a phenomenological model for high-$T_c$ superconductors. The authors[12] found that phase fluctuations can account for several features of the high-$T_c$ superconductors, among them the existence of a disorder-driven pseudo-gap state[11]. To demonstrate the possible relevance of our work, in the inset of Fig. 2b we plot the density of states of the system below (brown) and well above (green) the SIT. Below the SIT the density of states exhibits regular BCS-like superconducting behaviour, while above the transition the density of states exhibits a pseudo-gap feature due to the contribution of the superconducting correlations on the insulating side. For weaker disorder, this feature is only observed at lower density, which might correspond to the fact that the pseudo-gap is solely a feature of underdoped systems. We believe that incorporating phase fluctuations into a microscopic model for the high-$T_c$ superconductors will prove useful in explaining many of the experimental features of these systems.

## METHODS

### The model

The negative-$U$ Hubbard model is described by the hamiltonian:

$$H = \sum_{i,\sigma}(\varepsilon_i + h_\parallel \sigma)C^+_{i\sigma}C_{i\sigma} - t\sum_{<ij>,\sigma}\left(e^{i\phi_{ij}}C^+_{i\sigma}C_{j\sigma} + e^{-i\phi_{ij}}C^+_{j\sigma}C_{i\sigma}\right) - U\sum_i C^+_{i\uparrow}C_{i\uparrow}C^+_{i\downarrow}C_{i\downarrow} \quad (1)$$

where $C_{i\sigma}$ and $C^+_{i\sigma}$ destroy and create an electron with spin $\sigma$ at site $i$, respectively. The first term describes the random potential on the two-dimensional lattice, with a possible Zeeman field $h_\parallel$, while the second one describes the hopping between nearest-neighbour sites. The phases $\phi_{ij}$ account for the orbital effects of the magnetic field. The last term describes the attractive interaction between electrons on the same site and is responsible for the emergence of superconductivity. All energies are expressed in units of $t$, the hopping matrix element. The system is characterized by the relative strength of the attractive interaction $U$ (taken to be $U = 2$ throughout the calculations) and disorder $W$, comprising fluctuations in the on-site energies $\varepsilon_i$, the parallel magnetic field $h_\parallel$, the average electron density $n$ and the perpendicular magnetic field $B$ ($B$ is characterized by the magnetic flux per square in units of the quantum flux $\phi_0 = hc/e$, where $h$ is Planck's constant, $c$ is the speed of light and $e$ is the charge on the electron

The partition function for this model is given by

$$Z = \iint D\{C_i, C^+_i\}\exp\left(-\int_0^\beta d\tau\left[\sum_{i\sigma}C^+_{i\sigma}(\tau)(-\partial_\tau + \varepsilon_i + h_\parallel\sigma)C_{i\sigma}(\tau) - \sum_{<ij>\sigma}(t_{ij}C^+_{i\sigma}(\tau)C_{j\sigma}(\tau) + c.c.) - U\sum_i C^+_{i\uparrow}(\tau)C^+_{i\downarrow}(\tau)C_{i\downarrow}(\tau)C_{i\uparrow}(\tau)\right]\right) \quad (2)$$

where $\beta = 1/k_B T$, with $k_B$ the Boltzman constant, and *c.c.* denotes complex conjugate. Applying a Hubbard–Stratonovic transformation, with $\Delta_i$ the local Hubbard–Stratonovic field, with amplitude $|\Delta_i|$ and phase $\theta_i$, the partition function becomes:

$$Z = \iint D\{\Delta_i, \theta_i\}D\{C_i, C^+_i\}\exp\left(-\int_0^\beta d\tau\left[\sum_{i\sigma}C^+_{i\sigma}(\tau)(-\partial_\tau + \varepsilon_i + h_\parallel\sigma)C_{i\sigma}(\tau) - \sum_{<ij>\sigma}(t_{ij}C^+_{i\sigma}(\tau)C_{j\sigma}(\tau) + c.c.) - \sum_i(\Delta_i(\tau)e^{-i\theta_i(\tau)}C^+_{i\uparrow}(\tau)C^+_{i\downarrow}(\tau) + c.c) + \sum_i\frac{|\Delta_i(\tau)|^2}{U}\right]\right) \quad (3)$$

**The Bogoliubov–de Gennes approximation**

The partition function can be evaluated in the saddle-point approximation. Then the effective hamiltonian

$$H_{BdG} = \sum_{i,\sigma}(\varepsilon_i + h\sigma)C^+_{i\sigma}C_{i\sigma} - t\sum_{<ij>,\sigma}\left(e^{i\phi_{ij}}C^+_{i\sigma}C_{j\sigma} + e^{-i\phi_{ij}}C^+_{j\sigma}C_{i\sigma}\right) + \sum_i\left(\Delta_i C^+_{i\uparrow}C^+_{i\downarrow} + H.c.\right) \quad (4)$$

where $\Delta_i$ are now constants that obey the self-consistent relation $\Delta_i = -U\langle C_{i\uparrow}C_{i\downarrow}\rangle$.

$H_{BdG}$ is diagonalized via a Bogoliubov transformation $\gamma_{n\sigma} = \sum_i\left(u_n(r_i)C^+_{i\sigma} + \sigma v_n(r_i)C_{i\bar\sigma}\right)$. This yields an equation for the local order parameter $\Delta_i$ in terms of the Bogoliubov amplitudes $u_n(i)$ and $v_n(i)$ [19],

$$\Delta_i = |U|\sum_n u_n(i)v^*_n(i). \quad (5)$$

$u_n(i)$ and $v_n(i)$ are determined from the BdG equations [19],

$$\begin{pmatrix} \hat\xi & \Delta_i \\ \Delta^*_i & -\hat\xi \end{pmatrix}\begin{pmatrix} u_n(i) \\ v_n(i) \end{pmatrix} = E_n\begin{pmatrix} u_n(i) \\ v_n(i) \end{pmatrix}, \quad (6)$$

where $\hat\xi$ is the single-particle part of the hamiltonian (5). Equations (5) and (6) are solved self-consistently to determine $\Delta_i$.

**Including phase fluctuations**

The Bogoliubov–de Gennes approximation completely neglects phase fluctuations of the order parameter, due to its mean-field nature. To account for thermal phase fluctuations, we ignore quantum fluctuations, that is, the time dependence of $\Delta$ in the partition function (3). The resulting partition function is:

$$Z = \int \prod_i d|\Delta_i|\,d\theta_i \exp\left(-\frac{\beta}{2U}\sum_i|\Delta_i|^2\right)\mathrm{Tr}\exp(-\beta H_{BdG}) \quad (7)$$

where $H_{BdG}$ is the Bogoliubov–de Gennes (BdG) hamiltonian (4), and so the partition function reads:

$$Z = \int \prod_i d|\Delta_i| d\theta_i \exp\left(-\frac{\beta}{2U}\sum_i |\Delta_i|^2\right)\prod_{n=1}^{2N}(1+\exp(-\beta E_n)) \tag{8}$$

where $E_n$ are the eigenvalues of $H_{BdG}$.

The evaluation of expectation values and correlation functions for this partition function is carried out numerically using a Monte Carlo scheme[19,20]: at each step, a set of values $\{|\Delta_i|,\theta_i\}_{i=1}^N$ is chosen, inserted into $H_{BdG}$, which is then diagonalized. The integrand of equation (8) is then evaluated and weighted with temperature. However, for low enough temperatures such that the Monte Carlo averages of $|\Delta_i|$ hardly differ from those obtained from their mean-field values, one may take $|\Delta_i|$ in equation (7) to be their mean-field values, and the integral runs over the phases only. The phase correlations $\langle\cos(\delta\theta_i - \delta\theta_j)\rangle$ are then evaluated by:

$$\langle\cos(\delta\theta_i - \delta\theta_j)\rangle = \frac{1}{Z}\int \prod_i d|\Delta_i| d\theta_i \cos(\delta\theta_i - \delta\theta_j)\exp\left(-\frac{\beta}{2U}\sum_i |\Delta_i|^2\right)\prod_{n=1}^{2N}(1+\exp(-\beta E_n)) \tag{9}$$

where at each Monte Carlo step only the phases $\theta_i$ are changed and each phase configuration is given its thermal weight according to equation (9).

## Acknowledgements

We acknowledge discussions with A. Auerbach. This work was carried out with the support of the Israel Science Foundation and the US-Israel binational science foundation. Y.D. acknowledges support from a Kreitman fellowship. Y.M. acknowledges the hospitality of the Aspen Center of Physics. Y.A. acknowledges JSPS fellowship.